# Optically confined polarized resonance Raman studies in identifying crystalline orientation of sub-diffraction limited AlGaN nanostructure

A.K. Sivadasan,[*] Avinash Patsha and Sandip Dhara[*]

Surface and Nanoscience Division, Indira Gandhi Centre for Atomic Research,

Kalpakkam-603102, India

## ABSTRACT

An optical characterization tool of Raman spectroscopy with extremely weak scattering cross section tool is not popular to analyze scattered signal from a single nanostructure in the sub-diffraction regime. In this regard, plasmonic assisted characterization tools are only relevant in spectroscopic studies of nanoscale object in the sub-diffraction limit. We have reported polarized resonance Raman spectroscopic (RRS) studies with strong electron-phonon coupling to understand the crystalline orientation of a single AlGaN nanowire of diameter ~100 nm. AlGaN nanowire is grown by chemical vapor deposition technique using the catalyst assisted vapor-liquid-solid process. The results are compared with the high resolution transmission electron microscopic analysis. As a matter of fact, optical confinement effect due to the dielectric contrast of nanowire with respect to that of surrounding media assisted with electron-phonon coupling of RRS are useful for the spectroscopic analysis in the sub-diffraction limit of 325 nm ($\lambda$/2N.A.) using an excitation wavelength ($\lambda$) of 325 nm and near ultraviolet 40X far field objective with a numerical aperture (N.A.) value of 0.50.

Email : sivankondazhy@gmail.com; dhara@igcar.gov.in

The research towards direct wide–band gap group III–nitride semiconducting materials and its alloys is very important because of its applications in blue to near UV light emitting devices.[1] Group III nitride materials are one of the most promising candidates for fabricating these short wavelength and high frequency optical devices.[2-4] InN, GaN, and AlN and its ternary (InGaN and AlGaN) and quaternary (InAlGaN) alloys are the well–known candidates for major optoelectronic applications.[3,5-7] The prominent candidate for this wide bang gap material is ternary alloy of $Al_xGa_{1-x}N$ system with the tunable band gap at 3.4–6.2 eV,[8,9] and is also an ideal candidate for the fabrication of such optoelectronic devices.[9,10] High crystalline and optical quality are required for the optoelectronic applications of these materials. Latest research in III-nitride achieved a great progress with a single nanowire light emitting diode (LED) source and photodetectors being integrated,[10] or optically coupled by waveguides on-chip.[11]

Since the size and shape of the nanowires affect the polarization anisotropy of optical absorption and emission spectra,[12] the polarized Raman scattering is known to be one of the most prominent local probes for understanding the lattice dynamics of the crystalline system and its crystallographic anisotropy.[13,14] An optical characterization tool of Raman spectroscopy with extremely weak scattering cross section tool is not popular to analyze scattered signal from a nanostructure in the sub-diffraction limit. In this regard, plasmonic assisted characterization tools, namely, surface and tip enhanced Raman spectroscopies are only relevant in spectroscopic studies of nanoscale object in the sub-diffraction limit.

In a far field optical configuration, we report the polarized resonance Raman spectra of a single AlGaN nanowire of diameter ~ 100 nm, using a monochromatic light source (λ) of 325 nm with a near ultraviolet (NUV) micro spot focusing objective lens (Thorlab-LMU-40X-NUV) of magnification 40X having numerical aperture (N.A.) of 0.50 and a working distance of 1mm. The focused spot diameter for 325 nm laser source can be found out as D = 1.22λ/N.A. = 793 nm (~ 0.8 μm). As per the Abbe diffraction limit (λ/2N.A.), it is not possible to focus an object below ~ 325 nm. The optical confinement of polarized light,[15] which may enhance the light-matter interaction between two semiconducting or dielectric media along with strong electron-phonon coupling in the polarized resonance Raman spectroscopy (RRS) are proposed to understand the crystallographic orientations in the sub-diffraction length scale.

Monodispersed AlGaN nanowires were synthesized on Si(100) substrates at 900 $^{o}$C by chemical vapour deposition technique in the Au catalyst assisted vapor-liquid-solid (VLS) process using Ga droplet (99.999%, Alfa Aesar), Al film (50 nm) as precursors and $NH_3$ (99.999%) with a flow rate 50 sccm as reactant gas. The Au catalysts were coated in the thermal evaporation technique (12A4D, HINDHIVAC, India) at with an evaporation time 5 min with a low evaporation rate of 1Å /sec and subsequent annealing at a temperature 900 $^{o}$C for 10 min. A two zone split furnace was used for the growth using Al precursor at 1100 $^{o}$C and Ga droplet at 900 $^{o}$C in the respective zones with same ramp rate of 15 $^{o}$C min$^{-1}$ for an optimized growth time of 120 min.

Morphological features of the as-prepared samples were analyzed using a field emission scanning electron microscope (FESEM, SUPRA 55 Zeiss). The structural and crystallographic nature of the AlGaN NWs were investigated with the help of a high resolution transmission electron microscopy (HRTEM, LIBRA 200FE Zeiss) by dispersing the NWs in isopropyl alcohol and transferred to Cu grids. The polarized vibrational properties of AlGaN NWs were studied using Raman spectroscopy (inVia, Renishaw, UK) with an excitation of 325 nm and 2400 gr.mm$^{-1}$ grating used as a monochromatizer for the scattered waves. The thermoelectrically cooled CCD detector was used in the backscattering geometry. The spectra were collected using a NUV 40X objective with numerical aperture (N.A.) value of 0.50.

Morphological features of mono-dispersed AlGaN nanowires are shown in the field emission scanning electron microscope image (Fig. 1). The high resolution FESEM image of the nanowires (inset Fig. 1) shows very smooth surface morphology with Au catalyst nanoparticles at the tip. Moreover, uniform size with an average diameter of 100(±10) nm was also observed for the nanowires. The well separated Au nanoparticles taking part in the VLS growth process of the nanowires were having uniform size of ~120 nm.

We investigated polarization dependent resonant Raman scattering experiments for a single AlGaN nanowire for understanding the crystalline orientation in the mono-dispersed sample. Since the Raman spectra obtained from a single nanowire by using 514.5 nm excitation shows very poor intensity (supplementary Fig. S1),[16] we performed RRS study using the 325 nm (3.81 eV) excitation above the band gap of wurtzite GaN (3.47 eV)[17] to maximize the Raman signal for a single nanowire in the sub-diffraction limit by invoking Fröhlich interaction of electron-phonon coupling.[18] In RRS, we excited the nanowires with a laser source having energy greater than the band gap of the material. So, it led to the interaction of conduction electron and longitudinal optical (LO) phonon of ionic crystals and hence an improvement of Raman intensities could be noticed. The electron-phonon coupling strength is estimated from the RRS studies of ensemble AlGaN (supplementary Fig. S2).[16] The polarized resonant Raman spectrum of single nanowire shows only one peak (Fig. 2(a)) centered around 725 cm$^{-1}$ which is assigned to $A_1$(LO) mode of the GaN. The peak observed at 521 cm$^{-1}$ is because of the optical vibrational mode of the Si(100) substrate. The sample is chosen for the polarized micro–Raman spectra in such a way that single crystalline AlGaN nanowire is horizontally laying on the Si(100) substrate itself. The axis along the cylindrical nanowire is chosen as X direction, where as the incident and the scattered light propagation direction is considered along the Z direction. The parallel $Z(XX\overline{Z}$ and perpendicular $Z(XY)\overline{Z}$ polarization was configured using a half wave plate and a polarizer.[14] In the defining the polarization configurations, the first and last letter represents the direction of the propagation of the incident light ($k_i$) and scattered light ($k_s$), respectively. The second and third letter inside the parenthesis represents the direction of the electric field polarization of the incident light ($E_i$) and scattered light ($E_s$), respectively. According to the polarization selection rule in the backscattering configuration for a wurtzite [0001] GaN, (i.e., Z∥[0001]), only the $E_2$(high) phonon mode should be observable in the $Z(XY)\overline{Z}$ while both $E_2$(high) and $A_1$(LO) phonon modes should be viewed in the $Z(XX\overline{Z}$ configuration.[14] However, it is well known that $E_2$(high) mode will not be highly active in the RRS. The $E_2$(high) mode, incidentally was observed for the 514.5 nm excitation wavelength (supplementary Fig. S1).[16] However, we preferred polarized RRS studies in these nanostructure as the $A_1$(LO) mode alone was detrimental in identifying the crystallographic information of wurtzite phase in the backscattering configuration.[14] Thus the $A_1$(LO) mode around 725 cm$^{-1}$, was observed in the $Z(XX\overline{Z}$ configuration and the same mode was absent in the case of the $Z(XY)\overline{Z}$ configuration. The schematic representation of the different configurations for polarized RRS and the possible growth orientation of AlGaN single nanowire are shown in the figure 2(b). Incidentally the optical mode of Si is also found to disappear in the $Z(XY)\overline{Z}$ configuration in the cubic symmetry.[19] Hence the presence of $A_1$(LO) mode in the $Z(XX\overline{Z}$ configuration and the absence of it in the $Z(XY)\overline{Z}$ configuration clearly indicates that the horizontally laying nanowires on the Si(100) substrate may have the crystallographic growth orientations normal to the nonpolar $m$–plane [1-100] (schematic Fig. 2(b)).

For further confirmation, a typical high resolution transmission electron micrograph is shown for a single nanowire (Fig. 3(a)). It shows an interplanar spacing of 2.98 Å corresponding to (1-100) *m*-planes of AlGaN. The SAED pattern of the nanowire (Fig. 3(b)) is indexed to wurtzite phase of single crystal AlGaN with zone axes along [0001]. So the growth direction of the AlGaN nanowire is along the *m*-planes of the wurtzite structure. The structural analysis is exactly matching with the information inferred from the polarized Raman study (Fig. 2). So the polarized Raman spectroscopy is an accurate tool for understanding the crystallographic orientations even at the nanometer scale. It was ensured that the individual nanowires are well separated (~ 3 μm as shown in inset in Fig. 2(a)) as compared to the excitation laser spot size ~ 0.8 μm to provide a Raman spectrum corresponding to a single nanowire. In the present study, the confinement of polarized light is because of the trapped photons in between semiconductor nanowire –air– semiconductor nanowire configuration as shown schematically (Fig. 4). Optical confinement of polarized light was reported in the 1D semiconductor-dielectric and metal-dielectric interfaces.[20-22] Since, there is a variation of refractive index in AlGaN nanowire–air–AlGaN nanowire interface, the excitation photons is trapped in between the columns of consecutive nanowires due to the multiple reflections from the surface of nanowires.[20,23] The optically trapped photons in between the consecutive nanowires interact with the material and yields the enhanced Raman spectra.[24,25] Use of RRS with strong electro-phonon coupling along with optical confinement of polarized light are also exploited to understand the crystallographic orientations for the sub-diffraction limit of ~100 nm nanowire using a wavelength of 325 nm. The optical confinement of polarized light in the nanostructures of diameter ~100 nm, which is well above the quantum confinement size of 11 nm for GaN,[26] is envisaged for the polarization measurements in the dielectric contrast of ~ 5.6 for AlGaN with respect to that of 1 for the surrounding medium of air.[27,28] It may be noted here that the diameter of the nanowire is far below the beam spot size of ~0.8 μm. However, the optical confinement of polarized light assisted with RRS have helped us to measure the intensities of allowed Raman modes in different polarization conditions to determine the crystalline orientation of a nanowire in the sub-diffraction limit.

In conclusion, mono-dispersed ternary AlGaN nanowires are grown using chemical vapour deposition technique following vapor-liquid-solid growth mechanism. Polarized resonance Raman spectroscopy studies with strong electron-phonon coupling is used to understand the crystalline orientation of a single AlGaN nanowire of diameter ~100 nm. The crystallographic orientations predicted by the polarized Raman spectroscopy shows a good agreement with the structural analysis. As a matter of fact, optical confinement effect due to the dielectric contrast of nanowire with respect to that of surrounding media along with strong electron-phonon coupling of resonance Raman spectroscopy is useful for the spectroscopic analysis in the sub-diffraction limit of ~100 nm using a wavelength of 325 nm.

One of us (AKS) acknowledges the Department of Atomic Energy for continuing the research work. We also thank A. Das, K. K. Madapu and Venkatramana Bonu, of SND, IGCAR for their valuable suggestions and useful discussions.

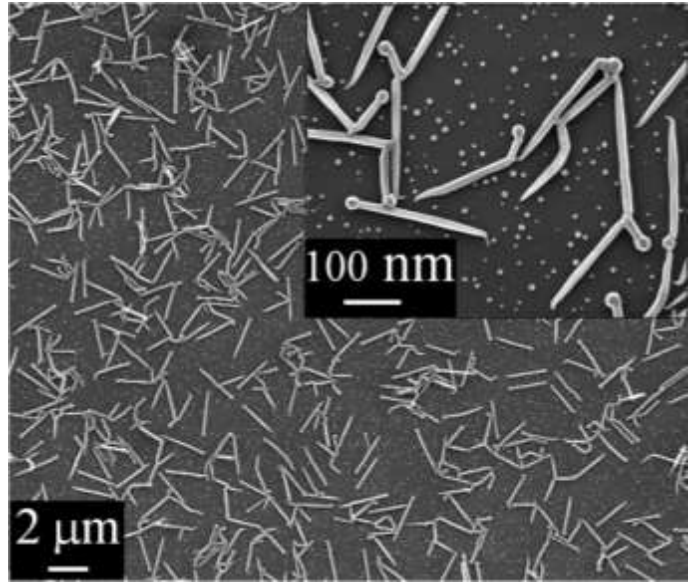

**FIG. 1.** FESEM image of AlGaN nanowires. Inset show high resolution images of the cylindrically shaped nanowire with Au nanoparticles at the tip.

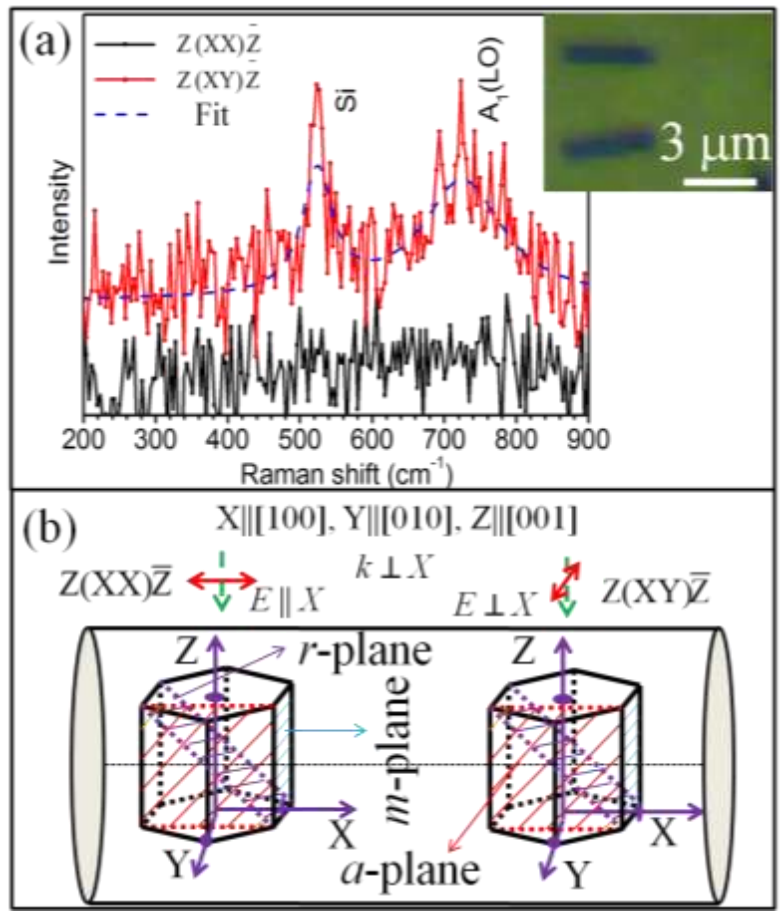

**FIG. 2.** (a) Polarized Raman spectra for single nanowire with an excitation wavelength 325nm for different polarization configurations. The AlGaN nanowire used for the Raman measurement is shown in the inset. (b) The schematic depiction for single nanowire with two different polarization configurations and the possible stacking arrangement of unit cells in the cylindrically shaped nanowire.

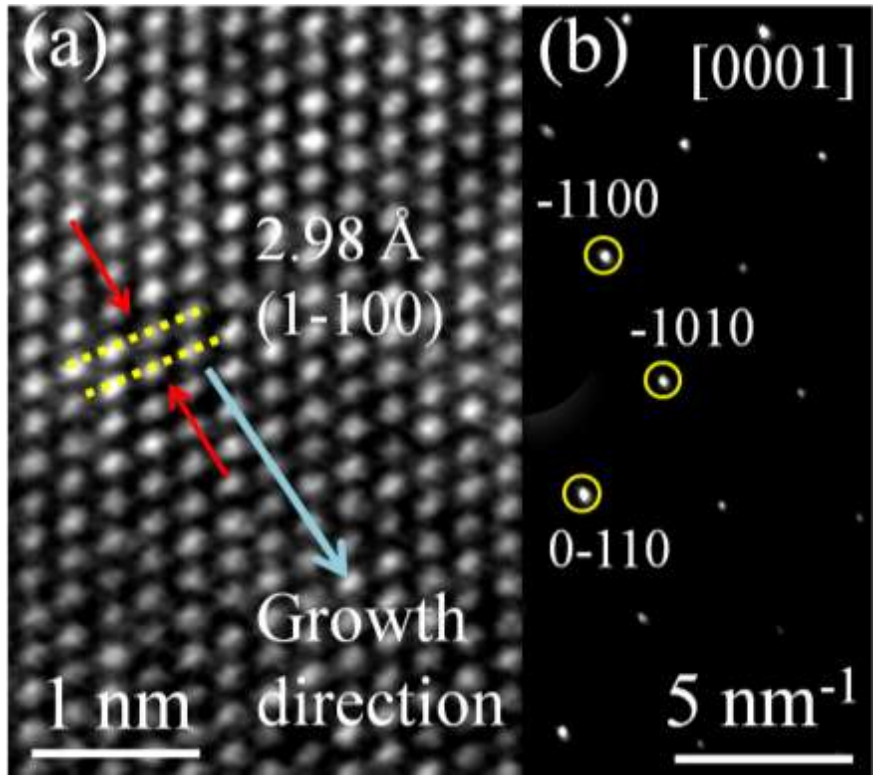

**FIG. 3.** (a) HRTEM image of the AlGaN nanowire shows an interplanar spacing of 2.98 Å and the growth direction along [1-100]. (b) SAED pattern of the nanowire indexed to the wurtzite phase of AlGaN with zone axis along [0001].

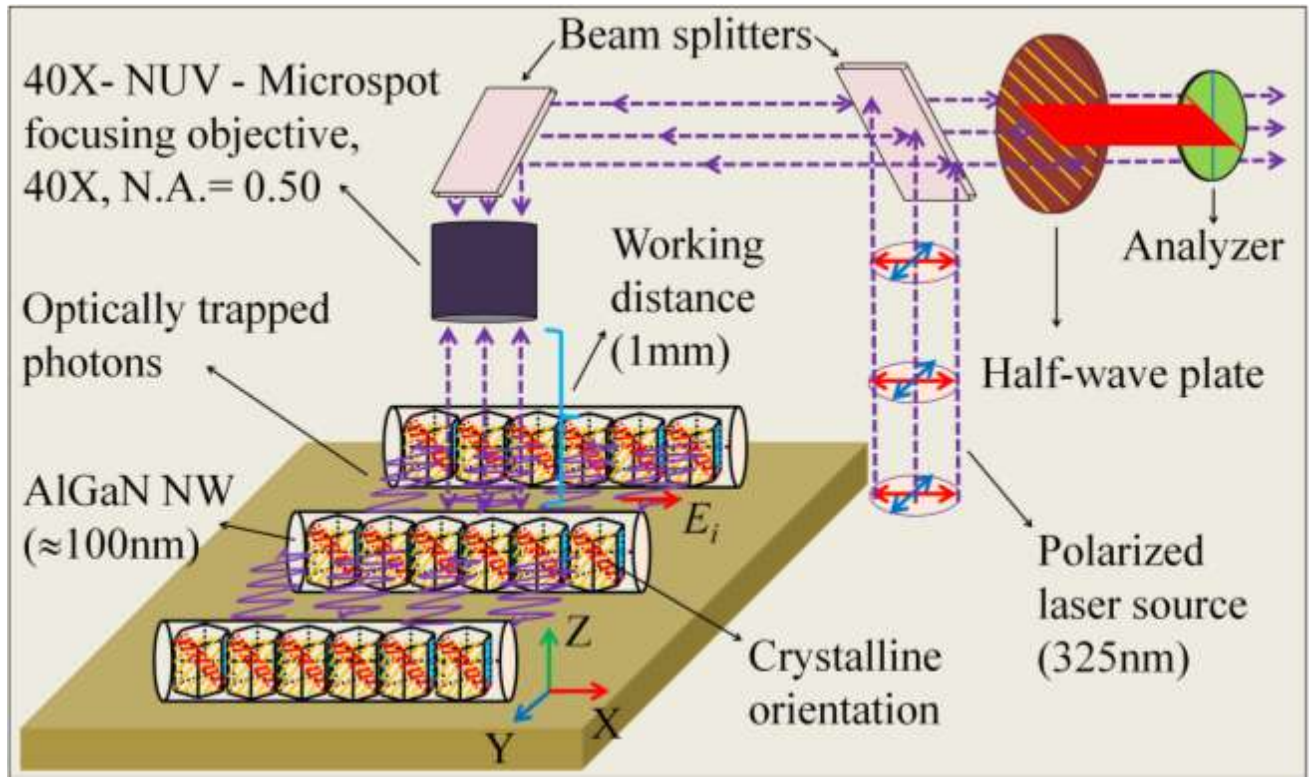

**FIG. 4.** The schematic representation of polarization configuration and optically trapped photons leading to the optical confinement effect in between the semiconducting nanowire columns. The plane waves are shown to bounce back and forth between the arrays of nanowires which may lead to the formation of standing waves. The confined electric vector ($E_i$) of the incident light is parallel to the long axis of the nanowire.